\documentclass{svjour2}

\usepackage{amsmath}
\usepackage{amssymb}
\usepackage{helvet}         
\usepackage{courier}        
\usepackage{type1cm}        
\usepackage{makeidx}         
\usepackage{graphicx}        
\usepackage{multicol}        
\usepackage[bottom]{footmisc}
\makeindex             

\begin{document}
	
	\def\beq{\begin{equation}}
	\def\endeq{\end{equation}}
	\def\begdi{\begin{displaymath}}
	\def\enddi{\end{displaymath}}
	\def\ep{\varepsilon}
	\def\speq{\hspace{1mm} = \hspace{1mm}}    
	\def\hilight{\textbf}
	\def\Z{\mathbb{Z}}
	\def\half{\frac{1}{2}}

\title{Creep tide theory. Equations for differentiated bodies with aligned layers}  
\author{Sylvio Ferraz-Mello, Hugo A.Folonier and Gabriel O. Gomes}
\institute{Instituto de Astronomia 
	Geof\'{\i}sica e Ci\^encias Atmosf\'ericas, Universidade de S\~ao Paulo, Brasil 
\email		sylvio [at] usp.br}
\maketitle

\abstract{The creep tide theory is used to establish the basic equations of the tidal evolution of differentiated bodies formed by aligned homogeneous layers in co-rotation. The mass concentration of the body is given by the fluid Love number $k_f$. The formulas are given by series expansions valid for high eccentricity systems. They are equivalent to Darwin's equations, but formally more compact. An application to the case of Enceladus, with $k_f=0.942$, is discussed.} 
	
\section{Introduction}

The creep tide theory proposed by Ferraz-Mello (2012, 2013) was initially established for the study of tidal friction in homogeneous bodies. It was later extended to bodies formed by concentric homogeneous layers by Folonier and Ferraz-Mello (2017). This paper aims at proposing a different extension of the theory to differentiated bodies able to give equations almost identical and simple as those of the homogeneous case, but valid for layered bodies. For that sake, we have to introduce several hypotheses, one of them being that all layers have the same angular rotation velocity and that their ellipsoidal boundaries have the same orientation. This new model is a reduced case of the general model proposed by Folonier and Ferraz-Mello (2017) where the layers are free to have different rotations and do not have their bounding ellipsoids with the same orientation. 
Because of the approximations being introduced, the aligned model is less precise than the general model proposed in 2017 but is adequate to study the tidal evolution in cases  in which the internal structure is totally unknown (as in the case of exoplanets) and condensed into just one free parameter, the fluid Love number.

Before proceeding with this objective, we remind that the creep tide theory considers that the body is a fluid and the tidal deformations are ruled by the Navier-Stokes equation for a low-Reynolds-number flow. The creep equation is an approximate solution of the hydrodynamical equations and the only free parameter is the viscosity of the fluid. No elastic terms are considered and there are no ad-hoc phase lags. The lags exist, but they are the characteristic forced lags appearing in the solution of first-order differential equations and are not free parameters.    

In its original formulation,
to get an analytical solution, the integration of the creep equation is iterative and separated into two steps. The first step, related to the instantaneous shape of the body, is initialized adopting a uniform rotation. The second step considers the variation of the orbital elements and spin. For a free rotating body, it was shown that the uniform rotation assumed in the first step is very close to the solution obtained at the second step. For bodies whose rotation is almost synchronous with the orbital motion, on the contrary, the solution shows an important oscillation (the forced libration) superimposed to the uniform rotation. In that case, the original formulation cannot be used and the best solution is then to adopt the parametric version of the creep tide theory (Folonier et al. 2018; Ferraz-Mello et al. 2020) in which the equations for the parameters defining the shape and orientation of the body, the orbital elements, and the spins, are integrated simultaneously. However, because of the short period variations, small integration steps are needed  and the parametric formulation cannot be used to investigate the long-term evolution of the system. Otherwise, it is necessary to improve the first iterative step of the original approach adding an undetermined oscillation to the uniform rotation (Folonier et al. 2018, Online Suppl.). The explicit consideration of the forced libration is necessary because it affects directly the dissipation of energy in the body.

In some sense, this paper is a revisit to the already published papers on the creep tide theory and an application of the equations obtained for the multi-layered model (Folonier and Ferraz-Mello, 2017) to the case of a differentiated body with a layered but rigid structure. The resulting equations are similar to those previously obtained for homogeneous bodies but introduce the factor $k_f$ as a measurement of the mass concentration. This factor is the same fluid Love number obtained from purely static models (Folonier et al. 2015; Wahl, 2017; Wahl et al. 2017). In addition, the paper includes  a coherent set of consolidated equations allowing an immediate application to actual problems in extra-solar systems.

Section 2 gives a summary of the full multi-layered model detailed and with all steps identified to allow its use. In section 3, the disturbing function of the full model is reduced to the case of frozen and aligned layers. In sections 4 to 6, they are given the variation equations for the orbital elements, for the rotation of the body, and for the dissipation. The last sections present
 a discussion of various points related to the results and an application to Enceladus. The conclusions are gathered in Section 9. At last, in an appendix, the equivalence of the creep tide theory and Darwin's theory for viscous bodies is considered. 

\section{The full multi-layered model}

Let us consider the disturbing potential at an external point $(r^*,\theta^*,\varphi^*)$, due to the non-sphericity of the i$^{\rm th}$ layer of one differentiated body of mass $m$ formed by $N$ concentric (but independent) homogeneous layers. The body is disturbed by one mass point $\tens{M}$ of mass $M$ orbiting at a distance $r$ from it\footnote{No hypothesis is done concerning the relative size of the two bodies. We may apply the given theory to any of the bodies of a 2-body system.}  
After the introduction of the Keplerian expansions, the k$^{\rm th}$ term of the resulting series is
\begin{eqnarray}
	\delta U_{ik}(r^*,\theta^*,\varphi^*) &=& -\displaystyle\frac{3GC_i\overline{\epsilon}_\rho E_{2,k}}{4r^{*3}}\sin^2{\theta^*}\frac{\Delta\big(R_i^5\mathcal{H}_{i}\cos{\sigma_{ik}}\cos{(2\varphi^*-\beta_{ik})}\big)}{\Delta(R_i^5)}\nonumber\\
	& & +\frac{G C_i\overline{\epsilon}_\rho E_{0,k}}{4r^{*3}}(3\cos^2{\theta^*}-1)\frac{\Delta\big(R_i^5\mathcal{H}_{i}\cos{\sigma''_{ik}}\cos{\beta''_{ik}}\big)}{\Delta(R_i^5)}\\ \nonumber
	& & + \delta_{0,k}  \frac{G C_i\overline{\epsilon}_z n^2}{2r^{*3}\Omega_N^2}(3\cos^2{\theta^*}-1)\frac{\Delta\big(R_i^5\mathcal{G}_i\cos{\sigma''_{ik}}\cos{\beta''_{ik}}\big)}{\Delta(R_i^5)},
\end{eqnarray}
where
\begin{itemize}
	\item 	$G$ is the gravitation constant;\\
	\item	$\Delta$ is the operator $\Delta(f_i)=f_i - f_{i-1}$ where $i-1$ and $i$ denote the inner and the outer boundary of the i$^{\rm th}$ layer, respectively; \\
	\item 	$R_i$ is the radius of the i$^{\rm th}$ boundary (outer boundary of the i$^{\rm th}$ layer); $R_0=0$; \\
	\item	$C_i$ is the polar moment of inertia of the i$^{\rm th}$ layer. In the spherical layers approximation: 
	\begin{equation}
		C_i \simeq \frac{8\pi}{15}d_i \Delta(R_i^5);   \label{eq:Ci}
\end{equation}
\item   $d_i$ is the density of the i$^{\rm th}$ layer; \\ 
	\item $\delta_{0,k}$ is the Kronecker delta ($\delta_{0,0}=1$ and $\delta_{0,k}=0$ when $k\neq0$);\\
	\item $\overline{\epsilon}_\rho$ is the mean flattening of the equivalent Jeans homogeneous 
	spheroid and $\overline{\epsilon}_z$ is the flattening of the equivalent MacLaurin homogeneous spheroid with the same rotation speed as the outermost layer:
	\begin{equation}
		\overline{\epsilon}_\rho = \frac{15MR_N^3}{4ma^3}; \ \ \ \ \ \ \ \ \ \ 
		\overline{\epsilon}_z = \frac{5\Omega_N^2R_N^3}{4Gm};
		\label{eq:ep}
	\end{equation}
		\item $E_{q,p}$ are the Cayley functions of the orbital eccentricity (Cayley, 1861)
	\begin{equation}
		E_{q,p}(e) = \frac{1}{2\pi}\int_0^{2\pi}\left(\frac{a}{r}\right)^3\cos{(qv+(p-q)\ell)\ d\ell};
		\label{eq:cayley}
	\end{equation}
	\item $\mathcal{H}_i$ and $\mathcal{G}_i$ (Clairaut's coefficients) are the functions of the internal structure introduced by Folonier et al. (2015): 
	\begin{equation}
		\mathcal{H}_i = \sum_{j=1}^N (\tens{E}^{-1})_{ij}x_j^3;\ \ \ \ \ \ \ \ \ \ \ \mathcal{G}_i = \sum_{j=1}^N (\tens{E}^{-1})_{ij}x_j^3\left(\frac{\Omega_j}{n}\right)^2;
		\label{eq:Clairaut-coe}
	\end{equation}
	\item $\Omega_i$ is the angular velocity of the of the i$^{\rm th}$ layer around an axis perpendicular to the orbital plane;\\
	\item   $x_i=R_i/R_N$ is the normalized mean equatorial radius\\
	\item $\displaystyle(\tens{E}^{-1})_{ij}$ are the elements of the inverse of the matrix $\tens{E}$, whose elements are
	\begin{equation}
		(\tens{E})_{ij} = \left\{\begin{array}{lll}\displaystyle - \frac{3}{2f_N}(\widehat{d}_j-\widehat{d}_{j+1})x^3_i,                                              & \  (i<j) \\
			\displaystyle - \frac{3}{2f_N}(\widehat{d}_i-\widehat{d}_{i+1})x^3_i + \frac{5}{2}- \frac{5}{2f_N}\sum_{k=i+1}^{N} (\widehat{d}_k-\widehat{d}_{k+1}) &(x_k^3-x_i^3), \\
			 \ \ \ \ &\ (i=j) \\
			\displaystyle - \frac{3}{2f_N}(\widehat{d}_j-\widehat{d}_{j+1})\frac{x^5_j}{x^2_i},                                             & \  (i>j) \end{array}\right.
	\end{equation}
	\item   $\widehat{d}_i=d_i/d_1$ is the normalized density of the i$^{\rm th}$ layer; $\widehat{d}_{N+1} \equiv 0$; \\ 
	\item   $f_N=\sum_{i=1}^N (\widehat{d}_i - \widehat{d}_{i+1}) x^3$ \\
	\item   $\sigma_{ik}$ and $\sigma''_{ik}$ are the phases defined by
	\begin{align}
		 \cos{\sigma_{ik}}   &= \frac{\gamma_i}{\sqrt{\gamma_i^2+(\nu_i+kn)^2}};   & \sin{\sigma_{ik}}   &= \frac{\nu_i+kn}{\sqrt{\gamma_i^2+(\nu_i+kn)^2}} \nonumber\\
		\cos{\sigma''_{ik}} &= \frac{\gamma_i}{\sqrt{\gamma_i^2+(kn)^2}};         & \sin{\sigma''_{ik}} &= \frac{kn}{\sqrt{\gamma_i^2+(kn)^2}};
		\label{eq:sigma_i}
	\end{align}
	\item $ \nu_i=2\Omega_i-2n $ is the semi-diurnal frequency of the i$^{\rm th}$ layer;\\ 
	\item $ \gamma_i$ is the relaxation factor at the boundary of the i$^{\rm th}$ and (i+1)$^{\rm th}$ layers;\\
	\item 	$\beta_{ik}$ and $\beta''_{ik}$ are the angles 
	\begin{eqnarray}
		\beta_{ik} = 2\varpi-(k-2)\ell+\sigma_{ik};\ \ \ \ \ \ \ \ \ \ \ 
		\beta''_{ik} = k\ell-\sigma''_{ik},
		\label{eq:delta_ik}
	\end{eqnarray}
	\item $a, e, n, \varpi, \ell, v$ are the semi-major axis, eccentricity, mean motion, longitude of the pericenter, mean and true anomalies.\\
\end{itemize}

Equation (1) formally differs from the corresponding equation in Folonier et al. (2017). The
constants $\mathcal{C}_{ik}, \mathcal{C}''_{ik}$ appearing in Eqn. (12) of Folonier et al. (2017) were substituted by their definitions
	\begin{eqnarray}
	\mathcal{C}_{ik} = \frac{1}{2}\mathcal{H}_i\overline{\epsilon}_\rho E_{2,k};\ \ \ \ \ \ \ \ \ \ \ \mathcal{C}''_{ik} = - \frac{1}{2}\mathcal{H}_i\overline{\epsilon}_\rho E_{0,k} 
			- \delta_{0,k}\frac{\mathcal{G}_i n^2 \overline{\epsilon}_z}{\Omega_N^2},
	\label{eq:C_ik}
\end{eqnarray}
where the previous definition of $\overline{\epsilon}_z$ has also been modified and now refers to the rotation of the outer layer instead of the synchronous rotation (see eqn. \ref{eq:ep}). One advantage of this formulation is to separate the tidal and rotational contributions of the zonal terms appearing mixed when
$\mathcal{C}''_{ik}$ is used. 

\section{Reduction to the aligned model}
In order to reduce the above model, we assume that all layers have their bulges aligned and the same rotation velocities, that is, for all $i$,
\begin{eqnarray}
	\Omega_{i} &\equiv & \Omega \nonumber \\
	\beta_{ik} &\equiv & \beta_k \nonumber 
\end{eqnarray}
which brings, as consequence,
\begin{eqnarray}
	\nu_i & \equiv & \nu \nonumber \\
	 \sigma_{ik} & \equiv & \sigma_k \nonumber \\
	\gamma_i & \equiv & \gamma \nonumber \\
	\sigma''_{ik} & \equiv & \sigma''_k \nonumber \\
	\beta''_{ik} & \equiv & \beta''_k \nonumber 
\end{eqnarray}
and
\begin{equation}
		 \mathcal{G}_i \speq  \mathcal{H}_i \left(\frac{\Omega}{n}\right)^2. \nonumber \\
\end{equation}
The new parameters introduced on the right-hand sides are defined by the given equations. The restrictions introduced by these  conditions will be discussed later (see Sec. \ref{final}).

The reduced form of the potential is
\begin{eqnarray}
	\delta U_{ik}(r^*,\theta^*,\varphi^*) &=& -\displaystyle\frac{3GC_i\overline{\epsilon}_\rho E_{2,k}}{4r^{*3}}\sin^2{\theta^*}\frac{\Delta(R_i^5\mathcal{H}_{i})\cos{\sigma_{k}}\cos{(2\varphi^*-\beta_{k})}}{\Delta(R_i^5)}\nonumber\\
	& & +\frac{GC_i\overline{\epsilon}_\rho E_{0,k}}{4r^{*3}}(3\cos^2{\theta^*}-1)
	\frac{\Delta(R_i^5\mathcal{H}_{i})\cos{\sigma''_{k}}\cos{\beta''_{k}}}{\Delta(R_i^5)}\\ \nonumber
	& & + \delta_{0,k}\frac{G C_i\overline{\epsilon}_z }{2 r^{*3}}(3\cos^2{\theta^*}-1)\frac{\Delta(R_i^5\mathcal{H}_i)\cos{\sigma''_{k}}\cos{\beta''_{k}}}
		{\Delta(R_i^5)},
\end{eqnarray}
or, making the sum over all layers:
\begin{eqnarray}
	\delta U_{k}(r^*,\theta^*,\varphi^*) &\speq& -\displaystyle\frac{Gk_{f}mR_N^2\overline{\epsilon}_\rho}{5r^{*3}}\sin^2{\theta^*}
	E_{2,k}\cos{\sigma_{k}}\cos{(2\varphi^*-\beta_{k})}\nonumber\\
	& & +\frac{Gk_{f}mR_N^2\overline{\epsilon}_\rho}{15r^{*3}}(3\cos^2{\theta^*}-1)E_{0,k}\cos{\sigma''_{k}}\cos{\beta''_{k}}
	\label{eq:deltaUk}
\nonumber\\
& &  + \delta_{0,k}\overline{\epsilon}_z \frac{2G  k_{f}mR_N^2}{15 r^{*3}}(3\cos^2{\theta^*}-1)\cos{\sigma''_{k}}\cos{\beta''_{k}},
\end{eqnarray}
where 
$k_{f}$ is the fluid Love number (Folonier et al. 2015): 
\begin{eqnarray}
	k_{f} = \frac{15}{4mR_N^2} \sum_{i=1}^N  \frac{\Delta(R_i^5\mathcal{H}_{i})C_i}{\Delta(R_i^5)}.  
\end{eqnarray}
or, considering Eqn. (\ref{eq:Ci}),
\begin{eqnarray}
	k_{f} = \frac{2\pi}{mR_N^2} \sum_{i=1}^N  R_i^5\mathcal{H}_{i}(d_i-d_{i+1}).
	\label{eq:k2}  
\end{eqnarray}
(N.B. $d_{N+1}=0$).
In the limit case of a homogeneous body, $k_f = 3/2$, and we obtain the same equations as in the original creep tide theory when zonal and tesseral contributions are both considered (Ferraz-Mello, 2015).

\section{Tidal perturbation in the semi-major axis and eccentricity}

In this section, we calculate the variations in semi-major axis and eccentricity due to tides raised in one of the bodies. (To have the complete picture, it is necessary to consider also the effects due to the tides raised in the other body). We follow the same procedures adopted in (Ferraz-Mello, 2013; 2015 and Folonier and Ferraz-Mello, 2017) to obtain the corresponding variational equations.

\subsection{The semi-major axis}
In this case, from Folonier and Ferraz-Mello (2017), we have:

\begin{eqnarray}
	\dot{a} &=& -\sum_{i=1}^N\frac{3C_in\overline{\epsilon}_\rho}{2ma}\sum_{k,j\in \mathbb{Z}}(k+j-2)E_{2,k}E_{2,k+j}\frac{\Delta\big(R_i^5\mathcal{H}_{i}\cos{\sigma_{ik}}\sin{(j\ell+\sigma_{ik})}\big)}{\Delta(R_i^5)}\nonumber\\
	& & -\sum_{i=1}^N\frac{C_in \overline{\epsilon}_\rho}{2ma}\sum_{k,j\in \mathbb{Z}}(k+j)E_{0,k}E_{0,k+j}\frac{\Delta\big(R_i^5\mathcal{H}_{i}\cos{\sigma''_{ik}}\sin{(j\ell+\sigma''_{ik})}\big)}
	{\Delta(R_i^5)}
\end{eqnarray}
or, in the reduction to the aligned layers case:
\begin{eqnarray}
	\dot{a} &=& -\frac{2k_f n R_N^2\overline{\epsilon}_\rho}{5a}\sum_{k,j\in \mathbb{Z}}(k+j-2)E_{2,k}E_{2,k+j}\cos{\sigma_{k}}\sin{(j\ell+\sigma_{k})}\nonumber\\
	& & -\frac{2k_f n R_N^2\overline{\epsilon}_\rho}{15 a}\sum_{k,j\in \mathbb{Z}}(k+j)E_{0,k}E_{0,k+j}{\cos{\sigma''_{k}}\sin{(j\ell+\sigma''_{k})}\big)}.	
	\label{eq:dot_a}
\end{eqnarray}

After the averaging over one orbital period, we obtain
\begin{eqnarray}
	\langle\dot{a}\rangle &=& -\frac{k_f n  R_N^2\overline{\epsilon}_\rho}{15a} 
	\sum_{k\in \mathbb{Z}}
	\left( 3(k-2) E_{2,k}^2\sin{2\sigma_{k}}
	+ k E_{0,k}^2 \sin{2\sigma''_{k}}\right)	
\end{eqnarray}

\subsection{Eccentricity}
In this case,

\begin{eqnarray}
&&	\dot{e} = -\sum_{i=1}^N\frac{3C_in\overline{\epsilon}_\rho}{2ma^2}\frac{(1-e^2)}{2e} \times \nonumber \\
	&& \left\{\sum_{k,j\in \mathbb{Z}}\Big(\frac{2}{\sqrt{1-e^2}}+(k+j-2)\Big)E_{2,k}E_{2,k+j}\frac{\Delta\big(R_i^5\mathcal{H}_{i}\cos{\sigma_{ik}}\sin{(j\ell+\sigma_{ik})}\big)}{\Delta(R_i^5) } \right. \nonumber\\
	& & \left.\qquad\qquad +\frac{1}{3}\sum_{k,j \in \mathbb{Z}}(k+j)E_{0,k}E_{0,k+j}\frac{\Delta\big(R_i^5\mathcal{H}_{i}\cos{\sigma''_{ik}}\sin{(j\ell+\sigma''_{ik})}\big)}
	{\Delta(R_i^5)} \right\}, 
\end{eqnarray}
or, after reduction to the aligned layers case,
\begin{eqnarray}
	\dot{e} &=& -\frac{3k_fnR_N^2\overline{\epsilon}_\rho}{15a^2}\frac{(1-e^2)}{e}\times \nonumber \\
	&& \qquad \left\{\sum_{k,j\in \mathbb{Z}}\Big(\frac{2}{\sqrt{1-e^2}}+(k+j-2)\Big)E_{2,k}E_{2,k+j}{\cos{\sigma_{k}}\sin{(j\ell+\sigma_{k})}\big)} \right.\nonumber\\
	& & \left. \qquad \qquad + \frac{1}{3} \sum_{k,j\in \mathbb{Z}}(k+j)E_{0,k}E_{0,k+j}{\cos{\sigma''_{k}}\sin{(j\ell+\sigma''_{k})}\big)}\right\}.
\end{eqnarray}

After the time-average over one orbital period, we obtain 

\begin{eqnarray}
	\langle \dot{e} \rangle &=&  -\frac{k_fnR_N^2\overline{\epsilon}_\rho}{30a^2 e } (1-e^2)	\nonumber\\ &&  \times \sum_{k\in \mathbb{Z}} \Big[
	3\Big(\frac{2}{\sqrt{1-e^2}}+(k-2)\Big)E_{2,k}^2\sin{2\sigma_{k}} 
	 + k E_{0,k}^2 \sin{2\sigma''_{k}} \Big].
	\label{eq:dot_e}
\end{eqnarray}

\section{Variation of the Rotation}

From Folonier and Ferraz-Mello (2017), the $z$-component of the torque acting on the body due to the tides raised on it by its companion is

\begin{eqnarray}
	M_{zi} = \frac{3GMC_i\overline{\epsilon}_\rho}{2a^3}\sum_{k,j\in \mathbb{Z}}
	E_{2,k}E_{2,k+j}\frac{\Delta\big(R_i^5\mathcal{H}_{i} 
		\cos{\sigma_{ik}}\sin{(j\ell+\sigma_{ik})}\big)}
	{\Delta(R_i^5)}.
	\label{eq:torque-creep}
\end{eqnarray}
or, making the sum over $i=1,N$ and introducing the fluid Love number,

\begin{eqnarray}
	M_{z} = \frac{2GMmR_N^2k_f\overline{\epsilon}_\rho}{5a^3}\sum_{k,j\in \mathbb{Z}}
	E_{2,k}E_{2,k+j}{ 
		\cos{\sigma_{k}}\sin{(j\ell+\sigma_{k})}\big)}.
\end{eqnarray}

After averaging over one orbital period:

\begin{eqnarray}
	\langle M_{z} \rangle = \frac{GMmR_N^2k_f\overline{\epsilon}_\rho}{5a^3}
	\sum_{k\in \mathbb{Z}} E_{2,k}^2  
		\sin{2\sigma_{k}}.
\end{eqnarray}

Hence

\begin{eqnarray}
	\langle \dot{\Omega} \rangle = - \frac{GMmR_N^2k_f\overline{\epsilon}_\rho}{5Ca^3}
	\sum_{k\in \mathbb{Z}} E_{2,k}^2  
	\sin{2\sigma_{k}}.
\end{eqnarray}

It is worth reminding that in the homogeneous case, $\displaystyle\frac{k_fmR_N^2}{C}=\frac{15}{4}$ and the above equation may be reduced to
\begin{eqnarray}
	\langle \dot{\Omega} \rangle = - \frac{3GM\overline{\epsilon}_\rho}{4a^3}
	\sum_{k\in \mathbb{Z}} E_{2,k}^2  
	\sin{2\sigma_{k}}.
\end{eqnarray}
(see Ferraz-Mello, 2013). 

In the general problem of the tidal perturbation, the study of the rotation of one differentiated body includes two other terms: the changes in the gravitational attraction of the layers due to their misalignment and the tidal friction due to the different rotational velocity of the layers (Folonier and Ferraz-Mello, 2017). These terms vanish in the case of a differentiated body  with aligned layers in co-rotation.   

\section{The dissipation}

The two only sources of the mechanical energy dissipated inside the planet are the changes of the orbital energy of the system and of the rotational energy of the considered body due to the tides raised on the body by a companion. 

Hence
\begin{equation}
\dot{W}=\dot{W}_{\rm orb} + \dot{W}_{\rm rot} =
 \frac{GMm}{2a^2}\dot{a}+C\Omega\dot{\Omega}.
\end{equation}

Obviously, we are assuming that the considered body is not active. The above consideration is not valid in the case of a star because stars are prone to have angular momentum leakages (Bouvier et al., 1997), and even mass losses that interfere in the energy balance. It is also not valid to apply this reasoning to individual layers. In the case of the total dissipation, the system of two bodies is a closed system and then the mechanical energy lost by the system may be equal to the energy dissipated. But this is not so for the individual layers.  

Since
\begin{eqnarray}
	\langle\dot{W}_{\rm orb}\rangle &=& - \frac{k_f GMm   R_N^2 n\overline{\epsilon}_\rho}{30a^3} \nonumber \\
	&& \qquad \times
	\sum_{k\in \mathbb{Z}}
	\left( 3(k-2) E_{2,k}^2\sin{2\sigma_{k}}
	+ k E_{0,k}^2 \sin{2\sigma''_{k}}\right)	
\end{eqnarray}
(keeping in $\langle\dot{W}_{\rm orb}\rangle$ only the part coming from the tides raised in the considered body) and
\begin{eqnarray}
	\langle \dot{W}_{\rm rot} \rangle = - \frac{k_f  GMm R_N^2\Omega\overline{\epsilon}_\rho}{5a^3}
	\sum_{k\in \mathbb{Z}} E_{2,k}^2  
	\sin{2\sigma_{k}}
\end{eqnarray}
there follows
\begin{eqnarray}\label{eq:Wdot}
	\langle\dot{W}\rangle &=& 
	-\frac{k_f GMm R_N^2\overline{\epsilon}_\rho}{30a^3} 
	\times  \\ &&
	\sum_{k\in \mathbb{Z}}
	\Big[ \big( 6  (\Omega-n) + 3kn  \big)  E_{2,k}^2\sin{2\sigma_{k}}	                 		
	+ kn E_{0,k}^2 \sin{2\sigma''_{k}}\Big].	\nonumber
	\end{eqnarray}

Strictly speaking, to have a complete picture, we should also consider the internal mechanical energy of the planet that  changes in the process as the tidal forces affect the semi-axes of the ellipsoid representing the planet. However, this is a small quantity oscillating about zero. The same may be said of the cross terms of energy variation related to the changes in one body due to the tidal deformations of the other. These terms exist, but they are of the order of the product of the flattenings of the two bodies and, thus, negligible in first-order theories. In these theories, the tides raised in one body only take into account the central term of the gravitational potential of its companion.      

\section{Discussion}\label{final}

One of the more stringent points in the aligned layers model is that the alignment implies that the relaxation factors are equal in all layers' boundaries. This is not expected. The radial creep of the contents of a layer is not isotropic and the equilibrium bulges may have any orientation, depending on the viscosities of the adjacent layers. This condition may however be circumvented, but the expressions for the resulting fluid Love number may become awkward (Folonier, in preparation).

Another point deserving to be highlighted is the reduced role played by the polar oblateness of the body. It does not affect the rotation and contributes only short-period terms in the variations of the semi-major axis and eccentricity of the system. One can see it by inspecting the expansion of the corresponding term in Eqn. \ref{eq:deltaUk}: 

\begin{eqnarray}
	\delta_z U_{0}(r^*,\theta^*,\varphi^*) &\speq& \overline{\epsilon}_z \frac{2G  k_{f}mR_N^2}{15 a^{3}}(3\cos^2{\theta^*}-1)	\sum_{j\in \mathbb{Z}}  E_{0,j}\cos{j\ell}.
\end{eqnarray}
This is the disturbing potential due to the oblateness of the body (not tidal) and the only long-term perturbations arising from it are a constant perturbation in the mean motion and a precession of the pericenter.

The restriction to aligned co-rotating layers can also be formally overcome. However, when this is done, the resulting generalized Love number is no longer just a function of the mass concentration of the body, but a quantity that also  depends on several other parameters describing the properties of the layers. In addition, the resulting fluid Love number becomes slightly time-dependent (Netelmann, 2019; Folonier, in preparation). Except for some bodies of our solar system, we do not have enough observational constraints to determine the differences implied by a more complex model.

Many versions of the Darwin theory use the formulas given by Love (1909) and consider ab initio the disturbing potential with a constant factor $k_2$ (or $k_f$) without giving an explicit form to $k_2$
(e.g. Jeffreys, 1961).
Indeed, at the usual order of approximation, the disturbing potential depends only on the moments of inertia around the principal axes and any expansion may be reduced to a form similar to that obtained for the aligned layers model with an undetermined $k_f$. 

\section{Application to Enceladus}

In this section, we report a short application of the above model to Enceladus. Enceladus became a sort of benchmark for tidal theories since classical theories were shown to be insufficient to explain the observed high heat dissipation of that satellite. We use in this application the model of internal structure proposed by Beuthe et al. (2016). With that model and Eqn. \ref{eq:k2}, we obtain\footnote{
We may compare the value thus obtained to the value 0.962 that we obtain using the improved Darwin-Radau approximation as given by Ragazzo (2020) and the \textit{Cassini} determination of the moment of inertia of Enceladus (Iess et al., 2014).} $k_f=0.942$. The dissipation $\dot W$ may be calculated with the lowest-order approximation of Eqn. \ref{eq:Wdot}:
\begin{equation}\label{Estat}
	[\langle \dot{W}_{\rm stat}\rangle] = -\frac{21k_{f}GM^2n R^5} {2a^6}{e^2}\frac{\gamma n}{\gamma^2+n^2}
	+ \mathcal{O}(e^4).
\end{equation} 
Adopting the public data available on Enceladus, the relaxation factor may be obtained from 
\begin{equation}
\gamma_N=\frac{d_N g_N R_N}{2\eta_N}
\end{equation}
(Folonier and Ferraz-Mello, 2017) and extended to the boundaries of all layers (they are equal in the aligned model). We also adopt, for the viscosity of the outermost layer ($\eta_N$), the same range of values adopted by Roberts and Nimmo (2008) for the viscosity of the ice shell ($10^{13}-10^{14}$ Pa s) (the geometric mean of this range is $3.2 \times 10^{13}$ Pa s, very close to the $2.4 \times 10^{13}$ Pa s found by Efroimsky (2018a)). Hence, we obtain for the  dissipation power of
Enceladus, the (geometric) mean value 11 GW, which may be compared 
to the estimations
of the heat dissipated in the anomalously warm south polar terrain (SPT) which radiates up to 15.8 GW as indicated by thermal infrared Cassini data (Howett et al. 2011).

This application is just an example. In a complete study, it is  necessary to consider the forced oscillation of the satellite around the stationary rotation (forced libration) which influences the amount of energy dissipated in the body and may be responsible
for a 27\% increase in the dissipation of Enceladus (Folonier and Ferraz-Mello, 2018; Efroimsky 2018b).

\section{Conclusion}

The conclusion is that the creep tide equations for a differentiated body formed by aligned co-rotating layers are the same as for a homogeneous body with only a different fluid Love number. 
Hence, the previous results for homogeneous bodies are extensible to differentiated bodies just replacing, in front of each equation, one factor 3/2 (fluid Love number of homogeneous bodies) by the actual fluid Love number $k_f$ defined by Eqn.
\ref{eq:k2}.

The variation of the elements and rotation presented in this paper are the same for which explicit series expansions to the order ${\cal O}(e^2)$ were given in (Ferraz-Mello, 2022). The full expansions, with the Cayley coefficients determined by their integral definition, are preferable in applications to exoplanetary systems since approximations to ${\cal O}(e^2)$ are only valid for low eccentricities.

As an example of application, the model has been used to estimate the fluid Love number of Enceladus, from a set of physical parameters related to the inner structure, and to estimate the resulting tidal dissipation. The result ($\sim$11 GW) is in good agreement with the Cassini observations. The origin of the low dissipation value obtained with the classical  models by several authors 
  may be traced back to the use of Kelvin's formula (Thomson, 1863) for $k_2$ (see Efroimsky, 2015) and arbitrarily fixed values for the
  rigidity leading to $k_2 \le 0.002$.
  
The equivalence of the creep tide theory to the original Darwin's theory for viscous bodies, beyond the restriction to the cases where $\gamma \gg n$, is discussed in the Appendix.
The said restriction only appears when the creep tide theory is compared to the so-called Darwin's CTL theories and is due to the stringent hypotheses adopted in the CTL theories. These restrictions do not exist in the actual Darwin theory (Darwin, 1879, 1880) and the two theories give equivalent results.

\begin{acknowledgement}
	We thank the reviewers for their comments and suggestions. This investigation is sponsored by 
	CNPq (Proc. 303540/2020-6) and FAPESP (Procs. 2016/13750-6 ref. PLATO mission, 2016/20189-9 and 2017/25224-0).
	
\end{acknowledgement}

\section*{Appendix: Equivalence of Darwin's tidal theory for viscous bodies and the creep tide theory}

In our previous papers, the equivalence of the variational equations derived from the creep tide theory and those of Darwin's constant time lag (or CTL) theories was several times stressed. 
This equivalence is reinforced in this paper by the extension of the creep tide theory to a differentiated body with aligned corotating layers and the introduction of the actual fluid Love numbers\footnote{The Love number $k_f$ appearing in the first papers on the creep tide theory (Ferraz-Mello 2012, 2013, 2015) was just the Love number of fluid homogeneous bodies, $k_f=1.5$, and cannot be considered as an actual use of the fluid Love numbers}.

The restriction to the CTL theories stems from the fact that all versions of Darwin's theory published in the XX$^{th}$ century (revisited in Ferraz-Mello et al. 2008) followed what was dubbed ``Fall schwacher Reibung" by Gerstenkorn (1955), or ``weak friction approximation" (Alexander, 1973), in which the phase shifts, or lags, $\sigma_k$ are assumed to be small quantities. 
This postulate introduces in the theories one stringent approximation: Darwin's ``height" (also called ``fraction of equilibrium tide") $\cos\sigma_k$ becomes, in the first order of approximation, equal to 1 and disappears from the equations. 
When the factors $\cos\sigma_k$ missing in the CTL theories are reintroduced, we have total equivalence of the creep tide theory and Darwin's theory for homogeneous bodies.

The approach resulting from the introduction of the weak friction hypotheses was discussed by Efroimsky and Williams (2009) in a section of their paper, with the title ``The stone rejected by the builders". 
They showed that the weak friction approach was the culprit for some apparent singularities appearing in the equations, near the synchronism of rotational and orbital motions, when the ad hoc lags were taken proportional to a negative power of the frequency. 

In (Darwin, 1880a), the  phase shifts are inserted by hand, both in the case where they are kept undetermined and in the cases where they are fixed in agreement with his 1879 paper. However, 
under no circumstances did he assume that the phase shifts are small. On the contrary, there are in his paper (Darwin, 1880a) examples with phase shifts close to 45 degrees (the angles $f, g, \dots$ adopted by him correspond to $\half \sigma_k$ in the creep tide theory). High phase shifts appear in the tidal deformation of bodies with very high viscosity (see examples in Ferraz-Mello et al., 2020).

Differences however exist between the two theories. The most obvious one is that the quantities playing the role of relaxation factor in these theories, $\mathfrak{p}$ and $\gamma$, respectively, are related to the viscosity according to different laws (such that $\gamma = 4.75 \mathfrak{p}$). In addition, even in the case of viscous bodies, Darwin prefers to introduce the phase shifts by hand while in the creep tide theory they are introduced through the (approximated) solution of the creep differential equation. Major differences, however, appear when we consider the parametric version of the creep tide theory (Folonier et al, 2018; Ferraz-Mello et al. 2020). These equations, allow us to obtain a system of differential equations for the parameters defining the shape, orientation, and rotation of the body. The simultaneous integration of these equations can be done without the need for  any hypotheses on the rotation of the deformed body.

\end{document}